\documentstyle[12pt]{article}
\input{psfig}
\newcommand{\simleq}
{\mbox{\raisebox{-0.5ex}{$\textstyle \sim$}
 \raisebox{ 0.8ex}{$\textstyle  \!\!\!\!\!\! <$  }}}
\newcommand{\AmS}{{\protect\the\textfont2
  A\kern-.1667em\lower.5ex\hbox{M}\kern-.125emS}}

\begin{document}
\title{
Thermodynamical instability of
self-gravitational heavy neutrino
matter\footnote{To
be published in the Proceedings of 2nd
International Conference on Dark Matter in
Astro and Particle Physics (DARK98),
Heidelberg, Germany, 20-25 July 1998.}
 }
\author{
Neven Bili\'c$^1$
and
Raoul D.~Viollier$^2$
 \\
 $^1$Rudjer Bo\v{s}kovi\'{c} Institute, 10000 Zagreb, Croatia \\
$^2$Department of Physics,
University of Cape Town, \\ Rondebosch 7701, South Africa
}
\date{\today}
\maketitle
%-----------------------------------------------------------------------
\begin{abstract}
It is shown, in the framework of the Thomas-Fermi model
at finite temperature, that a cooling non-degenerate gas of massive
neutrinos will, at a certain  temperature, become unstable and undergo
a first-order phase transition in which quasi-degenerate supermassive
neutrino stars are formed through gravitational collapse.
For neutrinos in the mass range of 10 to 25 keV/c$^{2}$, these compact
dark objects could mimic the role of supermassive black holes that are
reported to exist at the centres of galaxies and quasi-stellar objects.
\end{abstract}

\section{Introduction}
A gas of massive fermions, interacting only gravitationally, has
interesting thermal properties that may have important consequences in
astrophysics and cosmology.  The canonical and grand-canonical ensembles for such a
system have been shown to have a non-trivial thermodynamical
limit \cite{Thi1,Her2}.  Under certain conditions, these systems
will become unstable and undergo a phase transition that is accompanied by
gravitational collapse \cite{Mes3}.  It is interesting to
note that this phase transition occurs only in the case of the attractive
gravitational interaction of neutral particles obeying Fermi-Dirac
statistics: it neither happens in the case of the repulsive
Coulomb interaction of charged fermions \cite{Fey4}, nor does it in the
case of the attractive gravitational interaction, when the
particles obey Bose-Einstein or Maxwell-Boltzmann statistics.  To
be specific, we henceforth assume that this neutral fermion is
the heaviest neutrino, which is presumably the $\nu_{\tau}$,
although this is not essential for most of the subsequent
discussion, as this neutral fermion could also be replaced with
a light gravitino in supergravity models with low supersymmetry breaking
scales \cite{Dicus}.
We will, moreover, assume that the weak interaction can
be neglected.

The ground state of a gravitationally condensed neutrino cloud,
with mass below the Oppenheimer-Volkoff (OV) limit \cite{RDV9},
where it is close
to being a black hole, is a cold neutrino star
\cite{Lur5,RDV6,RDV7,RDV8}, in which
the degeneracy pressure balances the gravitational attraction of
the neutrinos.  Degenerate stars of neutrinos in the mass range
between $m_{\nu}$ = 10 and 25 keV$/c^{2}$ are particularly
interesting \cite{RDV7,RDV8,RDV9}, as they could explain,
without resorting to the
black hole hypothesis, at least some of the features observed
around the supermassive compact dark objects, which are reported to
exist at the centres of a number of galaxies
\cite{Ton9,Dre10,Dre11,Kor12,Kor13,Kor14}, including
our own \cite{Lac15,Lac16,Eck26,Gen27,Tsi28,Ghez34,Tsik35}, and quasi-stellar
objects (QSO) \cite{Lyn17,Zel18,Bla19,Beg20,Tsi29}.  In fact, the
upper bound for the mass of a neutrino star is given by the
OV limit $M_{{\rm OV}} = 0.54195 m_{Pl}^{3}
m_{\nu}^{-2} g_{\nu}^{-1/2}$, where $m_{Pl} = (\hbar c/G)^{1/2}$
= 1.22205 $\cdot 10^{19}$ GeV/$c^{2}$ is the Planck mass and
$g_{\nu}$ the spin degeneracy factor of the neutrinos and
antineutrinos (i.e.\ $g_{\nu} = 2$ for Majorana or $g_{\nu} = 4$
for Dirac neutrinos and antineutrinos).  The radius of such a
compact dark object would be $R_{{\rm OV}} = 4.4466 R_{{\rm
OV}}^{s}$, where $R_{{\rm OV}}^{s} = 2GM_{{\rm OV}}/c^{2}$ is the
Schwarzschild radius of the mass $M_{{\rm OV}}$ \cite{RDV9}.  There is thus
little difference between a neutrino star at the
OV limit and a supermassive black hole of
the same mass, a few Schwarzschild radii away from the object, as
the last stable orbit around a black hole has a radius of 3 Schwarzschild
radii anyway.
For non-relativistic neutrino stars that are well below the
OV limit, mass and radius scale as $MR^{3} =
91.869 \hbar^{6} G^{-3} m_{\nu}^{-8} (2/g_{\nu})^{2}$
\cite{RDV8}. For instance, the purported supermassive black hole Sgr A$^{*}$
at the centre of our galaxy could well be a neutrino
star with 2.6 $\times$ 10$^{6}$ solar masses and a few tens of light-days
radius \cite{Tsi28,RDV9,RDV10,Tsik35}.

The existence of a quasi-stable neutrino in this mass range is ruled out
neither by particle and nuclear physics experiments nor by direct
astrophysical observations \cite{RDV8}.  On the contrary, if the LSND
experiment is actually observing $\nu_{\mu} \rightarrow \nu_{e}$
oscillations \cite{Ath22}, and the quadratic see-saw mechanism involving
the up, charm and top quarks, is the correct explanation for the
smallness of the neutrino masses \cite{Gel23,Yan24}, we should expect the
$\nu_{\tau}$ mass between 6 keV/$c^{2}$ and 32 keV/$c^{2}$. In such a
scenario, the $\nu_{e}$ and $\nu_{\mu}$ masses would be about 10$^{-5}$
and 1eV, respectively, and the solar and atmospheric neutrino anomalies
would have to be interpreted as vacuum
oscillations into two different sterile neutrinos, $\nu_{e}^{S}$ and
$\nu_{\mu}^{S}$, with masses slightly different from those of $\nu_{e}$ and
$\nu_{\mu}$, consistent
with the measured solar and atmospheric neutrino mass squared differences.
Large neutrino-antineutrino asymmetries could suppress potentially disastrous
active-sterile neutrino oscillations prior to nucleosynthesis \cite{Foot1}.
The mixing angles for active-sterile neutrino oscillations involving the same
flavor would be maximal, while the mixing matrix elements corresponding to
flavor oscillations would be very small and exhibit a hierarchical pattern similar to
that of the CKM matrix of the quark sector. The smallness of the masses of the
sterile neutrinos $\nu_{e}^{S}$, $\nu_{\mu}^{S}$ and $\nu_{\tau}^{S}$ could be
guaranteed by an additional see-saw mechanism in the sterile sector
\cite{Berez}.

It is well known that a neutrino mass of between 10 keV/c$^{2}$ and
25 keV/c$^{2}$ which is in the so-called
cosmologically forbidden region given by
93 $h^{-2}$ eV/$c^{2} \; \simleq m_{\nu} \; \simleq$ 4 GeV/$c^{2}$, where
0.4 $\simleq h \; \simleq$ 1 is the Hubble parameter, is unacceptable, as
it would lead to an early neutrino-matter
dominated phase some time after nucleosynthesis and prior to
recombination \cite{Kol21}.  In such a universe, the microwave background
temperature would be attained much too early to accommodate the
oldest stars in globular clusters, nuclear cosmochronometry, and the
Hubble expansion age, if the Standard Model of Cosmology is
correct.  However, the early universe might have evolved quite
differently, in the presence of such a heavy neutrino.  In
particular, it is conceivable that primordial neutrino stars have
been formed in local condensation processes during a
gravitational phase transition that must have occurred some time
after the nonrelativistic heavy neutrinos started to dominate this
universe, a few weeks after the big bang. Aside from reheating
the gaseous phase of the heavy neutrinos, the latent heat
produced during the phase transition might have contributed
partly to reheating the radiation as well.  Moreover, the bulk
part of the heavy neutrinos (and antineutrinos) will have
annihilated efficiently into light neutrinos via the $Z^{0}$ in
the dense interior of these supermassive neutrino stars
\cite{Lur5,RDV6,RDV7,RDV8}.  In this context, it is interesting
to note that, even within the (homogeneous) Standard Model of
Cosmology, the reason why neutrino masses larger than 4
GeV$/c^{2}$ are again cosmologically allowed, is precisely that the
annihilation of the heavy neutrinos into light neutrinos reduces
the matter content of the universe.  Since both these processes,
reheating and annihilation, will increase the age of the
universe, or postpone the time when the universe reaches the
current microwave background temperature \cite{RDV8,Kol21}, it
does not seem excluded that a quasistable massive neutrino in the
mass range between 10 and 25 keV is compatible with cosmological
observations \cite{Lur5,RDV8}.

\section{The Thomas-Fermi model}
The purpose of this paper is to study the formation of such a neutrino
star as a consequence of a thermodynamical instability during a first-order
gravitational phase transition. For simplicity,
we assume that the neutrino star will be sufficiently below the
OV limit, so that we can treat this process
non-relativistically \cite{Bil25}.
The general-relativistic extension of the Thomas-Fermi model has been
discussed in \cite{Bil,RDV11}. The effects of general relativity become
important only
if the total rest-mass of the system is close to the
OV limit \cite{Bil,RDV9,RDV11}.
The gravitational potential $V(r)$
satisfies the Poisson equation
\begin{equation}
\Delta V   =   4 \pi G m_{\nu}^{2} n_{\nu},
\end{equation}
where the number density of the $\tau$ neutrinos (including antineutrinos) of
mass $m_{\nu}$ can be expressed in terms of the Fermi-Dirac distribution
at a finite temperature $T$ as
\begin{equation}
n_{\nu} (r)   =   \frac{g_{\nu}}{4 \pi^{2} \hbar^{3}} (2 m_{\nu} kT)^{3/2}
I_{\frac{1}{2}} \left( \frac{\mu - V(r)}{kT} \right).
\end{equation}
Here $I_{n} (\eta)$ is the Fermi function
\begin{equation}
I_{n} (\eta)   =   \int^{\infty}_{0} \frac{\xi^{n} d \xi}{1 + e^{\xi -
\eta}},
\end{equation}
and $\mu$ the chemical potential.  It is convenient to introduce
the normalized reduced potential
\begin{equation}
v   =   \frac{r}{m_{\nu} GM_{\odot}} ( \mu - V),
\end{equation}
$M_{\odot}$ being the solar mass, and the dimensionless variable $x =
r/R_{0}$ with the scale factor
\begin{eqnarray}
R_{0} & = & \left( \frac{3 \pi \hbar^{3}}{4 \sqrt{2} m^{4}_{\nu} g_{\nu} G^{3/2}
        M_{\odot}^{1/2}} \right)^{2/3} \nonumber \\[.3cm]
  & = & 2.1377\;\;{\rm lyr} \left( \frac{17.2\;\;{\rm keV}}{ m_{\nu}
        c^{2}} \right)^{8/3} g_{\nu}^{-2/3}.
\end{eqnarray}
Using equations (2) and (4), equation (1) takes the simple form
\begin{equation}
\frac{1}{x} \frac{d^{2} v}{d x^{2}} = - \frac{3}{2} \beta^{-3/2}
I_{\frac{1}{2}} \left( \beta \frac{v}{x} \right),
\end{equation}
where we have introduced the normalized inverse temperature $\beta =
T_{0}/T$, with $T_{0} = m_{\nu} GM_{\odot}/kR_{0}$.  In equation (6) we
recover, at zero temperature, the well-known Lan\'{e}-Emden differential
equation \cite{RDV6,RDV8}
\begin{equation}
\frac{d^{2} v}{dx^{2}} = - \frac{v^{3/2}}{\sqrt{x}}.
\end{equation}

The solution of the differential equation (6) requires boundary
conditions.  We assume here that the neutrino gas is enclosed in a
spherical cavity of radius $R$ corresponding to $x_{1} = R/R_{0}$.  We
further require the total neutrino mass to be $M_{\nu}$, and we
allow for the possibility of a pointlike mass $M_{B}$ at the origin,
which could be, e.g., a compact seed of baryonic matter.  $v(x)$ is
then related to its derivative at $x = x_{1}$ by
\begin{equation}
v' (x_{1}) = \frac{1}{x_{1}} \left( v(x_{1}) - \frac{M_{B} +
M_{\nu}}{M_{\odot}} \right),
\end{equation}
which, in turn, is related to the chemical potential by $\mu = k T_{0} v'
(x_{1})$.  $v(x)$ at $x=0$ is given by the point mass at the
origin, i.e.\ $M_{B}/M_{\odot} = v(0)$.

Similarly to the case of the Lan\'{e}-Emden equation, it is easy to show
that equation (6) has a scaling property: if $v(x)$ is a solution of
equation (6) at a temperature $T$ and a cavity radius $R$, then $\tilde{v}
(x) = A^{3}v(Ax)$, with $A > 0$, is also a solution at the
temperature $\tilde{T} = A^{4}T$ and the cavity radius $\tilde{R} = R/A$.

It is important to note that only those solutions that minimize the
free energy are physical.  The free-energy functional is defined
as \cite{Her2},
\begin{eqnarray}
F[n_{\nu}] = \mu[n_{\nu}] N_{\nu} - W[n_{\nu}] - kTg_{\nu} \int \frac{d^{3} rd^{3} p}{(2 \pi
         \hbar)^{3}} \times& & \nonumber
\end{eqnarray}
\begin{eqnarray}
{\rm ln} \left[ 1 + {\rm exp} \left( - \frac{p^{2}}{2 m_{\nu} kT} -
        \frac{V[n_{\nu}]}{kT} + \frac{\mu[n_{\nu}]}{kT} \right) \right], & &
\end{eqnarray}
where
\begin{equation}
V[n_{\nu}] = -Gm^{2}_{\nu} \int d^{3} r' \frac{n_{\nu}(r')}{|\vec{r} - \vec{r}'|},
\end{equation}
and
\begin{equation}
W[n_{\nu}] = - \frac{1}{2} Gm_{\nu}^{2} \int d^{3} rd^{3} r'
\frac{n_{\nu}(r) n_{\nu}(r')}{|\vec{r} - \vec{r}'|}.
\end{equation}
The chemical potential in equation (9) varies with density, so
that the number of neutrinos $N_{\nu} = M_{\nu}/m_{\nu}$ is kept
fixed.

All the relevant thermodynamical quantities, such as number
density, pressure, free energy, energy, and entropy, can be
expressed in terms of $v/x$, i.e.
\begin{equation}
n_{\nu} (x) = \frac{M_{\odot}}{m_{\nu} R_{0}^{3}} \frac{3}{8 \pi}
\beta^{-3/2} I_{\frac{1}{2}} \left( \beta \frac{v}{x} \right),
\end{equation}

\begin{equation}
P_{\nu}(x) = \frac{M_{\odot} T_{0}}{m_{\nu} R_{0}^{3} 4 \pi}
\beta^{-5/2} I_{\frac{3}{2}} \left( \beta \frac{v}{x} \right) =
\frac{2}{3} \epsilon_{{\rm kin}} (x),
\end{equation}

\begin{eqnarray}
F & = & \frac{1}{2} \mu N_{\nu} + \frac{1}{2} kT_{0} R_{0}^{3} \int d^{3}
        x n_{\nu}(x) \frac{v(x) - v(0)}{x} \nonumber \\
  & - & R_{0}^{3} \int d^{3} x P_{\nu}(x),
\end{eqnarray}

\begin{eqnarray}
E & = & \frac{1}{2} \mu N_{\nu} - \frac{1}{2} kT_{0} R_{0}^{3} \int
        d^{3}x n_{\nu}(x) \frac{v(x) + v(0)}{x} \nonumber \\
  &   & + R_{0}^{3} \int d^{3} x \epsilon_{{\rm kin}} (x),
\end{eqnarray}

\begin{equation}
S = \frac{1}{T} (E-F).
\end{equation}
\section{Numerical results}
We now turn to the numerical study of a system of self-gravitating
massive neutrinos with an arbitrarily chosen total mass $M = 10
M_{\odot}$, varying the cavity radius $R$. Owing to the scaling
properties, the system may be rescaled to any physically interesting
mass.  For definiteness, the $\nu_{\tau}$ mass is chosen as
$m_{\nu}$ = 17.2 keV$/c^{2}$ which is about the central value of the mass
region between 10 and 25 keV$/c^{2}$ \cite{RDV8}, that is interesting for our
scenario.  In Figure 1, our results for a gas of
neutrinos in a cavity of radius $R = 100 R_{0}$ are presented.  We find three
distinct solutions in the temperature interval $T = (0.049 \div 0.311)
T_{0}$; of these only two are physical solutions, namely, those for
which the free energy assumes a minimum.  The density distributions
corresponding to such two solutions are shown in the first plot in
Figure 1.

\begin{figure}
%\centerline{\psfig{file=fig1.ps,height=8.0cm}}
\vbox{\vskip230pt
\includegraphics{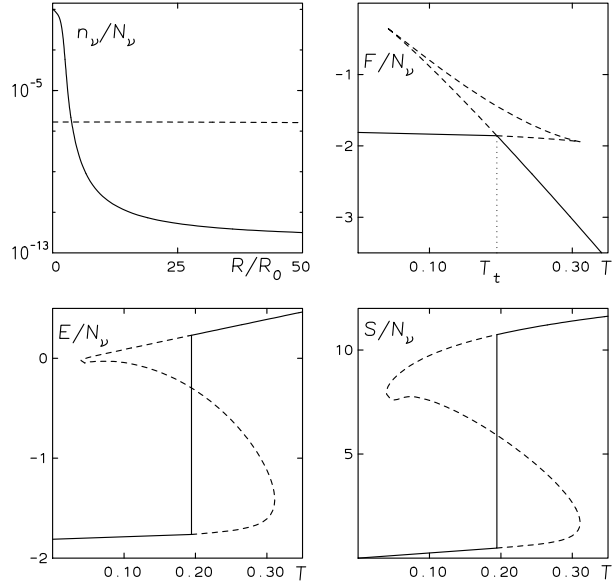}}
\caption{Density distribution normalized to unity for
condensate-like (solid line) and gas-like (dashed line) solutions
at $T=T_{t}$.  Free energy, energy, and entropy per particle as a
function of temperature.  Temperature, energy, and free energy
are in units of $T_{0}$.}
\label{fig1}
\end{figure}

The solution that can be continuously extended to any temperature
above the mentioned interval is referred to as ``gas'', whereas the
solution that continues to exist at low temperatures and
eventually becomes a degenerate Fermi gas at $T = 0$ is referred
to as ``condensate''.  In Figure 1, we also plot various
extensive thermodynamical quantities (per neutrino) as functions
of the neutrino temperature.  The phase transition takes place at
a temperature $T_{t}$, where the free energy of the gas and of the
condensate become equal.  The transition temperature $T_{t} =
0.19442 T_{0}$ is indicated by the dotted line in the free-energy
plot.  The top dashed curve in the same plot corresponds to the
unphysical solution.  At $T = T_{t}$ the energy and the entropy
exhibit a discontinuity, and thus there will be a substantial
release of latent heat during the phase transition.  An
important and currently still open question is, how and to which type of
matter or radiation this latent heat, which can be interpreted as
the binding energy of the neutrino stars, will be transferred.

We acknowledge useful discussions with Dr.\ D.\ Tsiklauri and
Dr.\ G.I.\ Kyrchev.

\end{document}